\documentclass[a4paper,11pt]{article}
\usepackage{here}
\usepackage{pos}
\usepackage{multirow}
\usepackage{slashed}
\usepackage{comment}

\title{Probing a light charged Higgs boson at the LHC Run 3}

\author[a]{A. Arhrib}
\author[b]{R. Benbrik}
\author*[c,\dag]{M. Krab}
\author[c]{B. Manaut}
\author[d,e]{S. Moretti}
\author[f]{Y. Wang}
\author[g,h]{Q. S. Yan}

\affiliation[a]{Abdelmalek Essaadi University, Faculty of Sciences and Techniques, B.P. 2117 Tétouan, Tanger, Morocco}

\affiliation[b]{Laboratoire de Physique Fondamentale et Appliquée de Safi, Faculté Polydisciplinaire de Safi, Sidi Bouzid, B.P. 4162, Safi, Morocco}

\affiliation[c]{Research Laboratory in Physics and Engineering Sciences, Modern and Applied Physics Team, Polydisciplinary Faculty, Beni Mellal, Morocco}

\affiliation[d]{School of Physics and Astronomy, University of Southampton, Southampton SO17 1BJ, UK}
\affiliation[e]{Department of Physics and Astronomy, Uppsala University, Box 516, SE-751 20 Uppsala, Sweden}
\affiliation[f]{College of Physics and Electronic Information, Inner Mongolia Normal University, Hohhot
	010022, P.R. China}
\affiliation[g]{Center for Future High Energy Physics, Chinese Academy of Sciences, Beijing 100049, P.R. China}
\affiliation[h]{School of Physics Sciences, University of Chinese Academy of Sciences, Beijing 100039, P.R. China}

\emailAdd{mohamed.krab@usms.ac.ma}

\note[\dag]{Address after October 27, 2023: Department of Physics, National Taiwan University}

\abstract{We study the Large Hadron Collider (LHC) discovery prospects of a light charged Higgs boson decaying into a $W$ boson and a non-Standard Model (SM)-like Higgs within the 2-Higgs Doublet Model type-I. 
In the analysis, we consider the associated production of a charged Higgs boson with a light neutral one, $pp \rightarrow H^{\pm} h$, with the subsequent $H^\pm \rightarrow W^{\pm*} h$. We then investigate the emerging  $W^{\pm*} + 4b$ final state and provide several benchmark points for signal-to-background analysis. We therefore  show that this signal could be an excellent avenue for identifying $H^\pm$ at the LHC.}

\FullConference{The European Physical Society Conference on High Energy Physics (EPS-HEP2023)\\
 21-25 August 2023\\
Hamburg, Germany\\}

\begin{document}
\maketitle

\section{Introduction}
In 2012, a scalar particle at the Large Hadron Collider (LHC) \cite{ATLAS:2012yve,CMS:2012qbp} with properties in agreement with the predicted Standard Model (SM) Higgs boson was discovered. The existence of the Higgs boson is therefore confirmed. 
But since then, no Beyond the SM (BSM) phenomena or particles have been seen at the LHC, which might revolutionise the present picture of particle physics. The SM, as it is known today, is the most widely accepted theory for describing elementary particles and their interactions.
However, there are several reasons (both theoretical and experimental) for believing that there is BSM physics.

The existence of charged Higgs bosons is a common prediction in  many BSM scenarios with an extended Higgs sector. Among these, the 2-Higgs Doublet Model (2HDM) is a well-motivated and simplest extension of the SM that agrees with the LHC Higgs boson discovery. The model is given by adding one additional scalar doublet to the SM. 
The interaction of the two Higgs doublets with SM fermions determines its well-known four types (type-I, type-II, type-X and type-Y). In the type-I realisation, which is the main focus of this study, only one doublet couples to fermions in the same way as the SM does.


In this work we analyse the LHC $W^{\pm*} + 4b$ signature\footnote{The $W^\pm + 4\gamma$ signature has been analysed in our previous works \cite{Wang:2021pxc,Wang:2021zjp}.} arising from the $H^\pm h$ charged Higgs production with its subsequent decay $H^\pm \rightarrow W^{\pm*} h$. We aim to provide Benchmark Points (BPs) and present a signal-to-background study at a center-of-mass energy of $14$ TeV and an integrated luminosity of $300~\rm{fb}^{-1}$. 

This paper is organised as follows. Section \ref{sect:2HDM} provides a brief review of the 2HDM. In Section \ref{sect:scan}, we describe the theoretical and experimental constraints to which the 2HDM is subjected. We discuss our analysis in Section \ref{sect:res} and we conclude in Section \ref{concl}.

\section{2HDM in a nutshell}
	\label{sect:2HDM}
The 2HDM is the simplest extension of the SM Higgs sector predicting the existence of a charged Higgs boson. It contains two complex $SU(2)$ doublet fields $\phi_1$ and $\phi_2$ with the same quantum numbers. The expression for the most general CP-conserving 2HDM with a softly broken $Z_2$ symmetry is as follows:
\begin{eqnarray}
V(\phi_1,\phi_2) &=& m_{11}^2(\phi_1^\dagger\phi_1) +
m_{22}^2(\phi_2^\dagger\phi_2) -
[ m_{12}^2(\Phi_1^\dagger\phi_2)+\text{h.c.}] ~\nonumber\\&& 
+\,\frac{1}{2}\lambda_1(\phi_1^\dagger\phi_1)^2 + \frac{1}{2}\lambda_2(\phi_2^\dagger\phi_2)^2 + 
\lambda_3(\phi_1^\dagger\phi_1)(\phi_2^\dagger\phi_2) ~\nonumber\\&&
+\,\lambda_4(\phi_1^\dagger\phi_2)(\phi_2^\dagger\phi_1) + \frac{1}{2}\lambda_5[(\phi_1^\dagger\phi_2)^2+\text{h.c.}],
\end{eqnarray}
where the parameters $m^2_{11}, m^2_{22}, m^2_{12}$ and $\lambda_{1-5}$ are real. The $Z_2$ symmetry is enforced to prevent the appearance of undesirable Flavour Changing Neutral Currents (FCNCs). The $m^2_{12}$ parameter allows for a soft violation of such a symmetry. 

Once Electro-Weak Symmetry Breaking (EWSB) occurs, three of the eight degrees of freedom initially present in the $SU(2)$ complex doublets ($\phi_1$ and $\phi_2$) are eaten up by the $W^\pm$ and $Z$ bosons, while the remaining five correspond to the physical Higgs fields $h$, $H$, $A$ and $H^\pm$. Consequently, the 2HDM Higgs sector includes three neutral Higgs states, two CP-even ($h$ and $H$ with $M_h < M_H$), and one CP-odd ($A$), as well as two charged Higgs states ($H^\pm$). One of the CP-even Higgs states should play the role of the SM-like one.

The 2HDM can be controlled by seven free parameters, namely $M_h$, $M_H$, $M_A$, $M_{H^\pm}$, $\tan\beta$, $\sin(\beta-\alpha)$ and $m^2_{12}$. The angle $\alpha$ is the mixing angle between the CP-even states ($h$, $H$) while $\tan\beta$ corresponds to the ratio of the Higgs doublets Vacuum Expectation Values (VEVs).  

\section{Scan methodology}
\label{sect:scan}
We built our analysis on a systematic scan of the 2HDM type-I parameter space using \texttt{2HDMC-1.8.0} \cite{Eriksson:2009ws}. The scan was carried out in our previous works \cite{Arhrib:2021xmc,Arhrib:2021yqf}, where we assumed that the discovered Higgs boson is the heavier CP-even Higgs $H$ with a mass of 125 GeV. 

During the scan, we required each sample point to satisfy unitarity \cite{Kanemura:1993hm}, perturbativity \cite{Chang:2015goa} and vacuum stability \cite{Barroso:2013awa} constraints. We checked the EW precision data through the oblique parameters $S$, $T$ and $U$ \cite{Grimus:2007if} using the best-fit results of Ref. \cite{Haller:2018nnx}. We considered exclusion limits from Higgs searches at collider experiments using \texttt{HiggsBounds-5} \cite{Bechtle:2020pkv} and ensured agreement with SM-like Higgs data using \texttt{HiggsSignals-2} \cite{Bechtle:2020uwn}. Additionally, we incorporated B-physics constraints using the results of Ref. \cite{Haller:2018nnx}. The 2HDM contribution to B-physics observables was evaluated using \texttt{SuperIso v4.1} \cite{Mahmoudi:2008tp}.

\begin{table}[H]
	\centering
	\begin{footnotesize}
		\begin{tabular}{|c| c| c| c| c| c| c|}
			\hline
			Parameters & BP1 & BP2 & BP3 & BP4 & BP5 & BP6 \\
			\hline
			$M_h$ & 65.11 & 69.88 & 69.12 & 64.39 & 65.20 & 68.65 \\
			\hline
			$M_A$ & 112.07 & 108.31 & 106.14 & 107.74 & 104.30 & 114.53  \\
			\hline
			$M_{H^\pm}$ & 88.51 & 85.50 & 90.62 & 107.61 & 106.02 & 115.66 \\
			\hline
			$\sin(\beta-\alpha)$ & $-0.061$ & $-0.059$ & $-0.092$ & $-0.059$ & $-0.064$ & $-0.098$ \\
			\hline
			$\tan\beta$ & 51.14 & 41.90 & 40.63 & 45.03 & 57.64 & 48.67 \\
			\hline
			$m^2_{12}$ & 82.33 & 113.63 & 115.73 & 90.47 & 73.50 & 96.16 \\
			\hline
			$\sigma(W^{\pm *}+4b)$ & 807.69 & 675.55 & 664.89 & 521.93 & 525.88 & 397.13 \\
			\hline
		\end{tabular}
		\caption{2HDM parameters and parton-level cross sections for the chosen BPs. All Higgs masses are given in GeV and for all BPs $M_H = 125$ GeV. Cross sections (in fb) are evaluated at $\sqrt{s} = 14$ TeV.}\label{t:bp}
	\end{footnotesize}
\end{table} 

In the feasible parameter space, which satisfies both the theoretical and experimental limits mentioned above, we select six BPs given in Table \ref{t:bp}. As seen in the table, the charged Higgs boson is light with mass between $85$ and $115$ GeV. Also, the neutral Higgs boson $h$ is always lighter than charged Higgs boson, kinematically enabling the $H^\pm \rightarrow W^\pm h$ decay.
Thus, the emerging $W^\pm$ boson will be off-shell since its mass is less than the mass difference between $H^\pm$ and $h$, i.e. $M_{H^\pm}-M_h < M_{W^\pm}$. Consequently, we anticipate soft charged leptons\footnote{In our analysis, we consider the leptonic decay of the $W$ boson, i.e. $W^\pm \rightarrow \ell^\pm \nu$ ($\ell = e, \mu$).}. 
In this study, our signal is given by $pp \rightarrow H^\pm h \rightarrow W^{\pm(*)} h h \rightarrow \ell^\pm \nu + 4b$ and the dominant SM background processes are $t\bar{t}$, $Wbbbb$, $Wbbjj$, $Wjjjj$ and $Ztb$, where $j$ corresponds to a light quark or gluon.

\section{Signal-to-background analysis}
\label{sect:res}
Before starting to discuss our analysis method, we present the toolbox used to generate both signal and background events. At the parton-level, these events are generated using the code \texttt{MadGraph5\_aMC@NLO-3.1.1} \cite{Alwall:2014hca}, where we adopt the two following kinematic cuts (in pseudorapidity, transverse
momentum, cone separation and Missing Transverse Energy (MET)):
\begin{align}
{\rm PC1:}\quad\quad
|\eta(\ell,{j})| < 2.5, \
p_T({j},\ell) > 10~\text{GeV}, \
\Delta R(\ell\ell/{jj}) > 0.4, \
\text{MET} > 5~\text{GeV},
\label{pc1}
\end{align}
\vspace*{-1.1truecm}
\begin{align}
{\rm PC2:}\quad\quad
|\eta(\ell,{j})| < 2.5, \
p_T({j},l) > 20~\text{GeV}, \
\Delta R(\ell\ell/{jj}) > 0.5, \
\text{MET} > 5~\text{GeV}.
\label{pc2}
\end{align}
The generating parton-level events are then passed to \texttt{Pythia-8.2} \cite{Sjostrand:2014zea} to include parton showering, hadronisation and hadron decays. For detector simulation, we utilise \texttt{Delphes-3.5.0} \cite{deFavereau:2013fsa} with the standard ATLAS card, where we adopt the \texttt{anti-kt} jet algorithm with a jet parameter $\Delta R = 0.5$. 

We first list in Table \ref{t:parton_cross_section} the cross sections of the signal and background processes after applying the parton-level cuts illustrated in eqs. \ref{pc1} and \ref{pc2}. The cross sections of our signal are around 20 to 32 fb for PC1, while they are between $2$ and $8$ fb for PC2.
  \begin{table}[H]
	\centering
		\begin{scriptsize}
			\begin{tabular}{|c|c|c|c|c|c|c||c|c|c|c|c|} 
				\hline
				$\sigma$ (fb) &  BP1&  BP2 &BP3 &BP4 &BP5 &BP6 & $t\bar{t}$& $Wbbbb$  & $Wjjbb$&  $wjjjj$ & $Ztb$\\
				\hline
				PC1 & 32.59 & 20.93& 26.22 &31.94 &31.38 & 26.40 & 85625&   9.45  & 13474   & 789960 & 0.143\\
				\hline
				PC2 & 5.39 & 2.71 &4.34 &8.31 &8.00  &7.89  & 54975  & 1.48 &2940 &  127545 & 9.3$\times 10^{-2}$ \\
				\hline
			\end{tabular}
		\end{scriptsize}
		\caption{The signal and backgrounds cross sections after the parton-level cuts.}\label{t:parton_cross_section}
\end{table} 

To reduce the backgrounds events, we apply the b-tagging. For this reason, we separate signal and background events into three categories: 4b0j (four $b$-jets, no light jets), 3b1j (three $b$-jets, one light jet) and 2b2j (two $b$-jets, two light jets).

 \begin{table}[H]
       \centering
		\begin{footnotesize}
			\begin{tabular}{|c|c|c|c|c|c|c|} 
				\hline
				& \multicolumn{3}{c|}{PC1}& \multicolumn{3}{c|}{PC2}\\
				\cline{2-7}
				&  4b0j & 3b1j & 2b2j &  4b0j & 3b1j & 2b2j\\
				\hline
				BP1 & 1.39 & 5.18 & 8.28 & 0.15 & 0.47 & 0.57 \\
				\hline
				BP2 & 0.86 & 3.03 & 4.71 & 0.08 & 0.21 & 0.26\\
				\hline
				BP3 & 1.16 & 4.20 & 6.64 & 0.13 & 0.38 & 0.47\\
				\hline
				BP4 & 1.78  & 6.34  & 10.22 & 0.31 & 1.01  & 1.28\\
				\hline
				BP5 & 1.74  & 6.18  & 9.83 & 0.31 & 0.95 & 1.21\\
				\hline
				BP6 & 1.67  & 5.72  & 9.03 & 0.34 & 0.99  & 1.26\\
				\hline\hline
				$t\bar{t}$ & 572.64 & 5226.43 & 29583.0 & 98.96 & 1658.4 & 14704.8 \\\hline
				$Wbbbb$ & 0.42 & 1.51 & 2.67 & 8.6$\times 10^{-2}$ & 2.61$\times 10^{-1}$ & 3.34$\times 10^{-1}$ \\\hline
				$Wjjbb$ & 36.69 & 354.22 & 2316.04 & 4.54 & 56.92 & 522.13 \\\hline
				$Wjjjj$ & 108.34 & 699.25 & 6480.41 & 6.96 & 89.81 & 939.82 \\\hline
				$Ztb$ & 0.022 & 0.054 & 0.073 & 9.53$\times 10^{-3}$ & 2.56$\times 10^{-2}$ & 3.02$\times 10^{-2}$\\
				\hline
			\end{tabular}
		\end{footnotesize}
		\caption{The signal and backgrounds cross sections after the pre-selection cuts.}\label{t:cross_section_bp23}
\end{table} 

In Table \ref{t:cross_section_bp23}, we list the cross sections of signal and background events in the three categories mentioned above. We observe that the cross sections are relatively small. This can be attributed to the circumstance that lepton reconstruction and b-tagging efficiencies depend on the transverse momenta of the involved objects. It can also be observed that the rates for PC1 are much larger than those for PC2, which is related to the fact that more leptons and b-jets (when $h$ is light) are soft. Thus, it is challenging for these soft objects to pass PC2, resulting in an event loss for PC2 when compared to PC1.

We turn now to compute the significance of our BPs. After the kinematic cuts, the significances are summarised in Table \ref{t:sig_bp20}. For PC1, we observe that almost all BPs possess a large significance for all categories. For PC2, the significance can be larger that 3 when the final state corresponds to the 4b0j category. In addition, we can obtain a sufficiently large significance in all cases by combining all the three final state categories. 

 \begin{table}[H]
	\centering
	\begin{footnotesize}
		\begin{tabular}{|c|c|c|c|c|c|c|c|} 
			\hline
			\multicolumn{2}{|c|}{} & BP1 & BP2 & BP3 & BP4 & BP5 & BP6 \\ \hline
			\multirow{3}{*}{PC1} & 2b2j & 3.65 & 2.19 & 3.01 & 3.56 & 3.55 & 2.85 \\\cline{2-8} & 3b1j & 8.51 & 5.10 & 6.82 & 8.12 & 7.96 & 6.41 \\\cline{2-8} & 4b0j & 8.79 & 6.06 & 7.21 & 9.08 & 9.43 & 7.74 \\\hline
			\multirow{3}{*}{PC2} & 2b2j & 0.45 & 0.27 & 0.51 & 0.73 & 0.71 & 0.70 \\\cline{2-8} & 3b1j & 1.60 & 1.30 & 1.90 & 2.97 & 2.42 & 2.37 \\\cline{2-8} & 4b0j & 3.28 & 2.45 & 3.30 & 5.44 & 4.91 & 4.79 \\\hline
		\end{tabular}
	\end{footnotesize}
	\caption{The significances corresponding to the selected BPs both for PC1 and PC2. Rates are evaluated at $\sqrt{s}$ = 14 TeV and $L = 300~\rm{fb}^{-1}$.}\label{t:sig_bp20}
\end{table} 

Finally, we refer to Ref. \cite{Li:2023btx} for further details and Refs. \cite{Arhrib:2022inj,Arhrib:2021yqf,Arhrib:2021xmc} for a general view of the 2HDM type-I parameter space.

\section{Conclusion}
\label{concl}
In the 2HDM type-I, we carried out a signal-to-background study for the EW $pp \rightarrow H^\pm h \rightarrow W^{\pm *} hh \rightarrow \ell^\pm \nu + 4b$ ($\ell=e,\mu$) process at the LHC. In the analysis, we included hard scattering, parton shower, hadronisation and detector effects using a set of tools, which allowed us to evaluate the possibility of observing the $\ell^\pm + 4b + \rm{MET}$ signal in a genuine experimental environment.

\section*{Acknowledgments}
The work of AA, RB, MK and BM is supported by the Moroccan Ministry of Higher Education and Scientific Research MESRSFC and CNRST Project PPR/2015/6.
SM is supported in part through the NExT Institute and STFC Consolidated Grant No. ST/L000296/1.
YW is supported by the Natural Science Foundation of
China Grant No. 12275143, the Inner Mongolia Science Foundation Grant No. 2020BS01013 and
Fundamental Research Funds for the Inner Mongolia Normal University Grant No. 2022JBQN080.
QSY is supported by the Natural Science Foundation of China Grant No. 12275143 and
No. 11875260.

\end{document}